\documentclass[proceedings]{JHEP3}
\usepackage{amsfonts}
\usepackage{amsmath}
\usepackage{epsfig,multicol}

\newbox\mybox

\newcommand\fverb{\setbox\mybox=\hbox\bgroup\verb}
\newcommand\fverbdo{\egroup\medskip\noindent\fbox{\unhbox\mybox}\ }
\newcommand\fverbit{\egroup\item[\fbox{\unhbox\mybox}]}
\conference{Breathers in the elliptic sine-Gordon model}
\abstract{We provide new expressions for the scattering amplitudes in the
soliton-antisoliton sector of  the elliptic sine-Gordon model in terms of
cosets of the affine Weyl group corresponding to
infinite products of q-deformed gamma functions. When relaxing the
usual restriction on the coupling constants, the model contains 
additional bound states
which admit an interpretation as breathers.
These breather bound states are unavoidably accompanied by Tachyons.
We compute the complete S-matrix describing the scattering of the
breathers amonst themselves and with the soliton-antisoliton sector.
We carry out various reductions of the model, one of them leading 
to a new type of theory, namely an elliptic 
version of the minimal $D_{n+1}^{(1)}$-affine Toda field theory.}

\title{Breathers in the elliptic sine-Gordon model}
\author{O.A.~Castro-Alvaredo and A.~Fring \\
Institut f\"ur Theoretische Physik, Freie Universit\"at Berlin, \\
Arnimallee 14, D-14195 Berlin, Germany \\
E-mail: \email{olalla/fring@physik.fu-berlin.de}}

\input{tcilatex}

\begin{document}

\section{Introduction}

By investigating a $\mathbb{Z}_{4}$-symmetry of the particle wavefunctions
for a soliton and an antisoliton, the elliptic sine-Gordon model was
introduced originally by A.B.~Zamolodchikov more than twenty years ago \cite%
{Z4}. The S-matrix was found to correspond to the transfer matrix of
Baxter's eight-vertex model \cite{Bax, Shankar:1981jq, Shankar:1982kz}. The
model has two free parameters $\nu $ and $\ell $ which were mutually
restricted in \cite{Z4} in order to avoid the presence of Tachyons. Thus, if
it was not for that restriction, the model could be viewed as a
generalization of the sine-Gordon model. Unfortunately, when constraining
mutually the parameters, it corresponds in the trigonometric limit only to
the soliton-antisoliton sector with no bound states. This means that the
entire breather sector is absent.

The purpose of this paper is to investigate whether a meaningful breather
sector of the model can be constructed. When relaxing the constraint on the
parameters several poles in the soliton-antisoliton S-matrix amplitudes will
move into the physical sheet. We will demonstrate below that some of them
lie on the imaginary axis and therefore, with the appropriate sign of the
residue, admit the usual interpretation as bound states corresponding to
breathers. In addition, there are redundant poles which are of a tachyonic
nature as they are in the physical sheet beyond the imaginary axis. In the
context of integrable models solutions for scattering amplitudes with such
properties have been discarded right away up to now. Nonetheless, Tachyons
have emerged initially undesired in other areas and subsequently have been
turned into virtues. For instance, recently there has been great activity in
the context of string theory \cite{Sen:1999nx, Berkovits:2000hf,
Kutasov:2000qp, Rastelli:2000hv, Harvey:2000na, Witten:2000nz, Adams:2001sv,
Andreev:2002sn, Sen:2002qa, Buchel:2002tj, Taylor:2002uv, Sen:2002an}, where
Tachyon condensates have turned out to be important. They also appear to be
very useful in the context of cosmological considerations \cite%
{Tomaschitz:1999tt, Choudhury:2002xu, Shiu:2002qe, Ohta:2002ac,
Gibbons:2002md, Kofman:2002rh, Gibbons:2003gb, Sami:2003qx, Kim:2003qz}.
Here we want to adopt the lesson from string theory and cosmology and allow
them to be present. The gain of this attitude is that we have additional
poles at our disposal which admit the usual interpretation as bound states,
which can be associated to breathers.

Our manuscript is organized as follows: In section 2 we assemble and derive
various properties of q-deformed gamma functions and relate them to Jacobian
elliptic functions. In section 3 we demonstrate how the affine Weyl group
can be utilized to solve the central functional equations leading directly
to an infinite product solution for the soliton-antisoliton backward
scattering amplitude in terms of q-deformed gamma functions. In section 4 we
discuss the soliton-antisoliton sector. Using a slightly modified procedure
proposed originally by Karowski and Thun \cite{Karo, KT}, the
soliton-breather amplitudes are constructed in section 5 and the breather
sector is discussed in section 6. By specifying the parameters to certain
values we reduce the model in section 7 to several models. Our conclusions
are stated in section 8.

\section{q-deformed gamma functions and Jacobian elliptic functions}

Many of the well studied integrable quantum field theories which allow for
backscattering, such as the sine-Gordon model, are known to possess a
generic form for their scattering amplitudes which consists of infinite
products of Euler's gamma functions 
\begin{equation}
S(\theta )=f(\theta )\prod\limits_{k=0}^{\infty }\prod\limits_{i=1}^{p}\frac{%
\Gamma (\theta +k\alpha +x_{i})}{\Gamma (\theta +k\alpha +y_{i})}\qquad
\quad \text{for }\sum\limits_{i=1}^{p}x_{i}=\sum\limits_{i=1}^{p}y_{i}~.
\label{gen}
\end{equation}
The rapidity dependent prefactor $f(\theta )$ is usually some finite product
of ratios of trigonometric functions and the values of $x_{i},y_{i},\alpha
,p $ are specific to each individual model. The constraint on the sums of
the $x_{i},y_{i}$ is a necessary condition for the convergence of the
infinite product (a proof of this can be found e.g.~in \cite{Migo}). Below
we will describe how the structure (\ref{gen}) can be generalized very
naturally by replacing in a controlled way the usual gamma functions by
their q-deformed counterparts and the trigonometric functions in the
prefactor by their elliptic version.

q-deformed quantities have turned out to be very useful objects as they
allow for instance to carry out elegantly (semi)-classical limits when the
deformation parameter is associated to Planck's constant. In the elliptic
sine-Gordon model we have the two free parameters $\nu \in \lbrack 0,\infty
) $ and $\ell \in \lbrack 0,1]$ at our disposal. The former is the analogue
of the coupling constant in the sine-Gordon model\footnote{%
This motivates to take $\nu $ to be positive, as negative values of it
correspond to a regime in which according to the arguments of Coleman \cite%
{Coleman} the ground state of the theory is not bounded from below.} and the
latter is the modulus of the Jacobian elliptic functions. It is the $\ell $
which is associated most naturally to a deformation. Accordingly, we define
a deformation parameter $q$ and its Jacobian imaginary transformed version,
i.e.~$\tau \rightarrow -1/\tau $, as 
\begin{equation}
q=\exp (i\pi \tau ),\qquad \hat{q}=\exp (-i\pi /\tau ),\qquad \tau
=iK_{1-\ell }/K_{\ell }~~.  \label{q}
\end{equation}%
We introduced here the quarter periods $K_{\ell }$ of the Jacobian elliptic
functions depending on the parameter $\ell \in \lbrack 0,1]$, defined in the
usual way through the complete elliptic integrals $K_{\ell }=\int_{0}^{\pi
/2}(1-\ell \sin ^{2}\theta )^{-1/2}d\theta \,$. Recalling the well-known
properties 
\begin{equation}
\lim_{\ell \rightarrow 0}K_{\ell }=\lim_{\ell \rightarrow 1}K_{1-\ell }=\pi
/2\qquad \text{and\qquad\ }\lim_{\ell \rightarrow 0}K_{1-\ell }=\lim_{\ell
\rightarrow 1}K_{\ell }\rightarrow \infty ~,  \label{kk}
\end{equation}%
the definitions (\ref{q}) obviously mean that the trigonometric limits
correspond to the \textquotedblleft classical\textquotedblright\ limit in
the variables $\hat{q},q$ as 
\begin{equation}
\lim_{\ell \rightarrow 0}\equiv \lim_{\hat{q}\rightarrow 1}\qquad \text{%
and\qquad }\lim_{\ell \rightarrow 1}\equiv \lim_{q\rightarrow 1}~.
\end{equation}%
It will turn out below that quantities in $\hat{q}$ will be most relevant
for our purposes and therefore we state several identities directly in $\hat{%
q}$, rather than $q$, even when they hold for generic values. The most basic
q-deformed objects one defines are q-deformed integers (numbers), for which
we take the convention 
\begin{equation}
\lbrack n]_{\hat{q}}:=\frac{\hat{q}^{n}-\hat{q}^{-n}}{\hat{q}-\hat{q}^{-1}}%
.\qquad \quad
\end{equation}%
They have the obvious properties 
\begin{eqnarray}
\lim_{\ell \rightarrow 0}[n]_{\hat{q}} &=&n,  \label{limn} \\
\lim_{\ell \rightarrow 0}\frac{[n+m\tau ]_{\hat{q}}}{[n^{\prime }+m^{\prime
}\tau ]_{\hat{q}}} &=&\left\{ 
\begin{array}{c}
1~~~~~~\qquad ~~\text{for~~~~}m,m^{\prime }\neq 0 \\ 
n/n^{\prime }\qquad ~~~~\ ~\text{for~~~~}m=m^{\prime }=0%
\end{array}%
\right. ~.  \label{limnn}
\end{eqnarray}%
With the motivation in mind mentioned at the beginning of this section we
define a q-deformed version of Euler's gamma function 
\begin{equation}
\Gamma _{\hat{q}}(x+1):=\prod\limits_{n=1}^{\infty }\frac{[1+n]_{\hat{q}%
}^{x}[n]_{\hat{q}}}{[x+n]_{\hat{q}}[n]_{\hat{q}}^{x}}~.\qquad  \label{gq}
\end{equation}%
The crucial property of the function $\Gamma _{\hat{q}}$, which coins also
its name, is 
\begin{equation}
\lim_{\ell \rightarrow 0}\Gamma _{\hat{q}}(x+1)=\lim_{\hat{q}\rightarrow
1}\Gamma _{\hat{q}}(x+1)=\prod\limits_{n=1}^{\infty }\frac{n}{n+x}\left( 
\frac{1+n}{n}\right) ^{x}=\Gamma (x+1)~.\qquad  \label{gqn}
\end{equation}%
We report now various properties of this function which will be useful
below. We can relate deformations in $q$ and $\hat{q}$ through 
\begin{equation}
\frac{\hat{q}^{(x+\tau /2-1/2)^{2}}}{\hat{q}^{(y+\tau /2-1/2)^{2}}}\frac{%
\Gamma _{\hat{q}}(y)\Gamma _{\hat{q}}(1-y)}{\Gamma _{\hat{q}}(x)\Gamma _{%
\hat{q}}(1-x)}=\frac{\Gamma _{q}(-y/\tau )\Gamma _{q}(1+y/\tau )}{\Gamma
_{q}(-x/\tau )\Gamma _{q}(1+x/\tau )}~.
\end{equation}%
Frequently we have to shift the argument by integer values 
\begin{equation}
\Gamma _{\hat{q}}(x+1)=\hat{q}^{x-1}[x]_{\hat{q}}\Gamma _{\hat{q}}(x)~.
\label{11}
\end{equation}%
Relation (\ref{11}) can be obtained directly from (\ref{gq}). As a
consequence of this we also have 
\begin{eqnarray}
\Gamma _{\hat{q}}(x+m) &=&\Gamma _{\hat{q}}(x)\prod\limits_{l=0}^{m-1}\hat{q}%
^{x+l-1}[x+l]_{\hat{q}}\quad \qquad \quad \quad ~~~~m\in \mathbb{Z}
\label{33} \\
\Gamma _{\hat{q}}(x) &=&\Gamma _{\hat{q}}(x-m)\prod\limits_{l=0}^{m-1}\hat{q}%
^{x-l-2}[x-l-1]_{\hat{q}}\quad \quad m\in \mathbb{Z~}.  \label{333}
\end{eqnarray}%
Whereas (\ref{11})-(\ref{33}) hold for generic $q$, the following identities
are only valid for $\hat{q}$ 
\begin{eqnarray}
\Gamma _{\hat{q}}(1/2-\tau /2)\Gamma _{\hat{q}}(1/2+\tau /2) &=&\ell
^{1/4}\Gamma _{\hat{q}}(1/2)^{2}  \label{2} \\
\frac{\Gamma _{\hat{q}}(x+2\tau )}{\Gamma _{\hat{q}}(y+2\tau )} &=&\frac{%
\Gamma _{\hat{q}}(x)}{\Gamma _{\hat{q}}(y)}  \label{ola1} \\
\prod\limits_{i=1}^{p}\frac{\Gamma _{\hat{q}}(x_{i})\Gamma _{\hat{q}%
}(x_{i}\pm \tau /2)}{\Gamma _{\hat{q}}(y_{i})\Gamma _{\hat{q}}(y_{i}\pm \tau
/2)} &=&\prod\limits_{i=1}^{p}\frac{\Gamma _{\hat{q}^{2}}(x_{i})}{\Gamma _{%
\hat{q}^{2}}(y_{i})}\quad ~~~\text{if\quad\ }\sum\limits_{i=1}^{p}x_{i}=\sum%
\limits_{i=1}^{p}y_{i}~  \label{4p} \\
\lim_{\hat{q}\rightarrow 1}\prod\limits_{i=1}^{p}\frac{\Gamma _{\hat{q}%
}(x_{i}\pm \tau /2)}{\Gamma _{\hat{q}}(y_{i}\pm \tau /2)} &=&1\quad \qquad
\qquad ~~~~~\text{if\quad\ }\sum\limits_{i=1}^{p}x_{i}=\sum%
\limits_{i=1}^{p}y_{i}~~~~  \label{5p} \\
\lim_{\hat{q}\rightarrow 1}\frac{1}{\ell ^{1/4}}\Gamma _{\hat{q}}(\frac{x}{%
2K_{\ell }}\mp \frac{\tau }{2})\Gamma _{\hat{q}}(1-\frac{x}{2K_{\ell }}\pm 
\frac{\tau }{2}) &=&\pi ~\qquad \qquad ~\text{for \quad ~~~}x\neq 0
\label{4}
\end{eqnarray}%
The constraint on the sums in (\ref{4p}) and (\ref{5p}) was already
encountered in (\ref{gen}). Most of these properties can be checked directly
by means of the defining relation (\ref{gq}). We comment on some of the
derivations below.

The singularity structure will be important for the physical applications.
It follows from (\ref{gq}) that the $\Gamma _{\hat{q}}$-function\ has no
zeros, but poles 
\begin{equation}
\lim_{\theta \rightarrow \theta _{\Gamma ,p}^{nm}=m\tau -n}\Gamma _{\hat{q}%
}(\theta +1)\rightarrow \infty \qquad \quad \text{for~~}~m\in \mathbb{Z}%
,n\in \mathbb{N}\mathbf{~.}  \label{pG}
\end{equation}
Furthermore, we will employ the Jacobian elliptic functions, to generalize
the prefactor $f(\theta )$ in (\ref{gen}), for which we use the common
notation $\func{pq}(z)$ with p,q $\in \{$s,c,d,n$\}$ (see e.g.~\cite%
{Elliptic} for standard properties). We derive important relations between
the q-deformed gamma functions and the Jacobian elliptic $\func{sn}$-function

\begin{eqnarray}
\func{sn}(x) &=&\frac{1}{\ell ^{\frac{1}{4}}}\frac{\Gamma _{\hat{q}}(\frac{x%
}{2K_{\ell }}\mp \frac{\tau }{2})\Gamma _{\hat{q}}(1-\frac{x}{2K_{\ell }}\pm 
\frac{\tau }{2})}{\Gamma _{\hat{q}}(\frac{x}{2K_{\ell }})\Gamma _{\hat{q}}(1-%
\frac{x}{2K_{\ell }})},  \label{sng} \\
&=&\frac{q^{\frac{1}{4}-\frac{ix}{2K_{1-\ell }}}}{i\ell ^{\frac{1}{4}}}\frac{%
\Gamma _{q}(\frac{1}{2}+\frac{ix}{2K_{1-\ell }})\Gamma _{q}(\frac{1}{2}-%
\frac{ix}{2K_{1-\ell }})}{\Gamma _{q}(1-\frac{ix}{2K_{1-\ell }})\Gamma _{q}(%
\frac{ix}{2K_{1-\ell }})}~.  \label{sng2}
\end{eqnarray}%
These relations can be used to obtain some of the above mentioned
expressions. For instance, recalling that $\func{sn}(K_{\ell })=1$, we
obtain (\ref{2}). With (\ref{gq}) we recover from this the well known
identity $\func{sn}(iK_{1-\ell }/2)=i/\ell ^{1/4}$. The trigonometric limits 
\begin{eqnarray}
\lim_{\ell \rightarrow 0}\func{sn}(x) &=&\lim_{\hat{q}\rightarrow 1}\func{sn}%
(x)=\frac{\pi }{\Gamma (\frac{x}{\pi })\Gamma (1-\frac{x}{\pi })}=\sin (x)
\label{ts} \\
\lim_{\ell \rightarrow 1}\func{sn}(x) &=&\lim_{q\rightarrow 1}\func{sn}(x)=%
\frac{1}{i}\frac{\Gamma (\frac{1}{2}+\frac{ix}{\pi })\Gamma (\frac{1}{2}-%
\frac{ix}{\pi })}{\Gamma (1-\frac{ix}{\pi })\Gamma (\frac{ix}{\pi })}=\tanh
(x).
\end{eqnarray}%
can be read off directly recalling (\ref{kk}), (\ref{gqn}) and presuming
that (\ref{4}) holds. We recall the zeros and poles of the Jacobian elliptic 
$\func{sn}(\theta )$-function, which in our conventions are located at 
\begin{eqnarray}
\text{zeros}\text{:\qquad } &&\theta _{\func{sn},0}^{lm}=2lK_{\ell
}+i2mK_{1-\ell }\qquad \quad \quad \quad l,m\in \mathbb{Z}  \label{zJ} \\
\text{poles}\text{:\qquad } &&\theta _{\func{sn},p}^{lm}=2lK_{\ell
}+i(2m+1)K_{1-\ell }\qquad ~l,m\in \mathbb{Z}\mathbf{~.}  \label{pJ}
\end{eqnarray}%
We have now assembled the main properties of the q-deformed functions which
we shall use below.

\section{Affine Weyl group and the unitarity/crossing relations}

The functional relations of crossing, unitarity and bootstrap for the
scattering amplitudes are usually solved by means of Fourier
transformations, thus leading in general directly to integral
representations for the desired quantities. When solving these integrals one
ends up with infinite products over gamma functions of the type (\ref{gen})
for scattering amplitudes in non-diagonal theories or trigonometric
functions when backscattering is absent. For the elliptic sine-Gordon model
so far only the analogue of the integral representation was presented \cite%
{Z4} in form of a discretized version. Instead of solving the discrete
integrals in this function we provide here a systematic procedure which
leads directly to product solutions by utilizing the affine Weyl group \^{W}$%
\left( \mathbf{g}\right) $.

Let us first assemble the necessary mathematical tools and jargon. In
general an affine Weyl reflection related to a simple root $\alpha _{i}$, of
a Lie algebra \textbf{g} \cite{Hum2} may be realized by the map 
\begin{equation}
\sigma _{i,n}\left( x\right) =x-\left( x\cdot \alpha _{i}\right) \alpha
_{i}+n\check{\alpha}_{i}~,  \label{affin}
\end{equation}
where $\check{\alpha}_{i}$ denotes a coroot and $n$ an arbitrary integer. In
particular for $n=0$ one recovers the ordinary Weyl group W$\left( \mathbf{g}%
\right) $ and for $x=0$ the translation group on the coroot lattice. Hence
the affine Weyl group may be thought of as a direct product

\begin{equation}
\hat{W}\left( \mathbf{g}\right) =W\left( \mathbf{g}\right) \otimes \mathbf{%
\check{T}~,}
\end{equation}
with $\mathbf{\check{T}}$ denoting translations on the coroot lattice. For
the case $\mathbf{g}=A_{1}~(\hat{W}\left( \mathbf{g}\right) \sim D_{\infty }$
the infinite dihedral group) one has only one simple root and (\ref{affin})
becomes 
\begin{equation}
\sigma _{n}\left( x\right) =n\alpha -x~.
\end{equation}
This group may be generated by two generators 
\begin{equation}
\sigma _{1}\left( x\right) =\alpha -x\qquad \hbox{and\qquad }\sigma
_{0}=-x\quad ,
\end{equation}
having obviously the property $\sigma _{1}^{2}=\sigma _{0}^{2}=1$. Defining
then the transformation

\begin{equation}
\sigma :=\sigma _{1}\sigma _{0}
\end{equation}%
one has 
\begin{equation}
\sigma ^{n}\left( x\right) =x+n\alpha \quad .
\end{equation}%
We denote by $N_{0},N_{1}$ the number of times the generators $\sigma
_{0},\sigma _{1}$ occur in an arbitrary element of $\hat{W}\left(
A_{1}\right) $. Then two types of subgroups $\left\{ \sigma
^{2n},N_{0}\right\} ,\left\{ \sigma ^{2n},N_{1}\right\} \subset \hat{W}%
\left( A_{1}\right) $ are constituted by the elements with $N_{0}$ and $%
N_{1} $ even, respectively. The right cosets of $\left\{ \sigma
^{2n},N_{1}\right\} $ may be divided into those with $N_{1}$ even $\left\{
\sigma ^{2n}I,N_{1}\right\} ,\left\{ \sigma ^{2n}\sigma _{0},N_{1}\right\} $
and $N_{1}$ odd $\left\{ \sigma ^{2n+1}I,N_{1}\right\} ,\left\{ \sigma
^{2n}\sigma _{1},N_{1}\right\} $. Similarly one may divide the right cosets
of $\left\{ \sigma ^{2n},N_{0}\right\} $ into those with $N_{0}$ even $%
\left\{ \sigma ^{2n}I,N_{0}\right\} ,\left\{ \sigma ^{2n}\sigma
_{1},N_{0}\right\} $ and $N_{0}$ odd $\left\{ \sigma ^{2n+1}I,N_{0}\right\}
,\left\{ \sigma ^{2n+1}\sigma _{1},N_{0}\right\} $.

Assuming now that $\hat{W}\left( A_{1}\right) $ acts in the complex rapidity
plane one may specify $\alpha $ and define the \textquotedblleft
unitarity\textquotedblright\ and \textquotedblleft
crossing\textquotedblright\ transformation on an arbitrary function $f\left(
\theta \right) $ by 
\begin{equation}
\sigma _{0}f\left( \theta \right) =f\left( -\theta \right) ,\qquad \sigma
_{1}f\left( \theta \right) =f\left( i\pi -\theta \right) ~.
\end{equation}
Then one obtains 
\begin{equation}
\sigma ^{n}f\left( \theta \right) =f\left( \theta +ni\pi \right) ~.
\end{equation}

We have now provided all the tools to solve the key equations in this
context. In \cite{Z4} two functional relations for the soliton-antisoliton
transmission amplitude (equations (2.10) and (3.8) therein), which we denote
by $c(\theta )$, were derived from crossing, unitarity and the Yang-Baxter
equations. Once this amplitude is known, some simple relations provided in 
\cite{Z4} suffice to construct the remaining ones in the soliton-antisoliton
sector. The equations to be solved are 
\begin{eqnarray}
c(\theta ) &=&c(i\pi -\theta )  \label{c1} \\
c(\theta )c(-\theta ) &=&\frac{\func{sn}^{2}(\pi /\nu )}{\func{sn}^{2}(\pi
/\nu )-\func{sn}^{2}(i\theta /\nu )}~.\quad \quad  \label{c3}
\end{eqnarray}%
We make now an ansatz by taking the ratio of right cosets in which $N_{0}$
is even and odd 
\begin{eqnarray}
c\left( \theta \right) &=&\kappa \frac{\left\{ \sigma ^{2n}I,N_{0}\right\}
\left\{ \sigma ^{2n}\sigma _{1},N_{0}\right\} }{\left\{ \sigma
^{2n+1}I,N_{0}\right\} \left\{ \sigma ^{2n+1}\sigma _{1},N_{0}\right\} }%
=\kappa \prod\limits_{k=1}^{\infty }\frac{\sigma ^{2k}\rho \left( \theta
\right) \sigma ^{2k}\sigma _{1}\rho \left( \theta \right) }{\sigma
^{2k+1}\rho \left( \theta \right) \sigma ^{2k+1}\sigma _{1}\rho \left(
\theta \right) },  \label{ansatz1} \\
&=&\kappa \prod\limits_{k=1}^{\infty }\frac{\rho \lbrack \theta +2\pi
ik]\rho \lbrack -\theta +2\pi i(k+1/2)]}{\rho \lbrack \theta +2\pi
i(k+1/2)]\rho \lbrack -\theta +2\pi i(k+1)]}~.  \label{ansatz}
\end{eqnarray}%
Here $\kappa $ is a constant and $\rho \left( \theta \right) $ an arbitrary
function which remains to be fixed. One observes that the crossing relation (%
\ref{c1}) is solved by construction, whereas (\ref{c3}) requires that 
\begin{equation*}
\kappa ^{2}\rho (\theta +2\pi i)\rho (2\pi i-\theta )=\frac{\Gamma _{\hat{q}%
}^{2}[-\frac{\tau }{2}]\Gamma _{\hat{q}}^{2}[1+\frac{\tau }{2}]\Gamma _{\hat{%
q}}[\hat{\theta}-\frac{\lambda }{2}]\Gamma _{\hat{q}}[1-\hat{\theta}+\frac{%
\lambda }{2}]\Gamma _{\hat{q}}[-\hat{\theta}-\frac{\lambda }{2}]\Gamma _{%
\hat{q}}[1+\hat{\theta}+\frac{\lambda }{2}]}{\Gamma _{\hat{q}}^{2}[-\frac{%
\lambda }{2}]\Gamma _{\hat{q}}^{2}[1+\frac{\lambda }{2}]\Gamma _{\hat{q}}[%
\hat{\theta}-\frac{\tau }{2}]\Gamma _{\hat{q}}[1-\hat{\theta}+\frac{\tau }{2}%
]\Gamma _{\hat{q}}[-\hat{\theta}-\frac{\tau }{2}]\Gamma _{\hat{q}}[1+\hat{%
\theta}+\frac{\tau }{2}]}.
\end{equation*}%
We replaced here in (\ref{c3}) the $\func{sn}$- by $\Gamma _{\hat{q}}$%
-functions using some standard identities for Jacobian elliptic functions
together with (\ref{sng}) and abbreviated for compactness 
\begin{equation}
\lambda =-\pi /K_{\ell }\nu \qquad \text{and\qquad }\hat{\theta}=i\theta
/2K_{\ell }\nu .
\end{equation}%
The problem of solving (\ref{c1}) and (\ref{c3}) has now been reduced to the
much simpler task of fixing $\rho \left( \theta \right) $ and $\kappa $. We
find 
\begin{equation}
\kappa =\frac{\Gamma _{\hat{q}}[-\frac{\tau }{2}]\Gamma _{\hat{q}}[1+\frac{%
\tau }{2}]}{\Gamma _{\hat{q}}[-\frac{\lambda }{2}]\Gamma _{\hat{q}}[1+\frac{%
\lambda }{2}]}\quad \quad \text{and\quad \quad }\rho _{i,j}(\theta )=\frac{%
\rho _{n,i}(\theta )}{\rho _{d,j}(\theta )}~,
\end{equation}%
where $\rho _{n,i}(\theta )$ and $\rho _{d,j}(\theta )$ could be any of the
functions 
\begin{equation*}
\begin{array}{ll}
\rho _{n,1}(\theta +2\pi i)=\Gamma _{\hat{q}}[\hat{\theta}-\frac{\lambda }{2}%
]\Gamma _{\hat{q}}[1-\hat{\theta}+\frac{\lambda }{2}]\quad & \rho
_{d,1}(\theta +2\pi i)=\Gamma _{\hat{q}}[\hat{\theta}-\frac{\tau }{2}]\Gamma
_{\hat{q}}[1-\hat{\theta}+\frac{\tau }{2}], \\ 
\rho _{n,2}(\theta +2\pi i)=\Gamma _{\hat{q}}[\hat{\theta}-\frac{\lambda }{2}%
]\Gamma _{\hat{q}}[1+\hat{\theta}+\frac{\lambda }{2}]\quad & \rho
_{d,2}(\theta +2\pi i)=\Gamma _{\hat{q}}[\hat{\theta}-\frac{\tau }{2}]\Gamma
_{\hat{q}}[1+\hat{\theta}+\frac{\tau }{2}], \\ 
\rho _{n,3}(\theta +2\pi i)=\Gamma _{\hat{q}}[-\hat{\theta}-\frac{\lambda }{2%
}]\Gamma _{\hat{q}}[1+\hat{\theta}+\frac{\lambda }{2}]\quad \quad & \rho
_{d,3}(\theta +2\pi i)=\Gamma _{\hat{q}}[-\hat{\theta}-\frac{\tau }{2}%
]\Gamma _{\hat{q}}[1+\hat{\theta}+\frac{\tau }{2}], \\ 
\rho _{n,4}(\theta +2\pi i)=\Gamma _{\hat{q}}[-\hat{\theta}-\frac{\lambda }{2%
}]\Gamma _{\hat{q}}[1-\hat{\theta}+\frac{\lambda }{2}] & \rho _{d,4}(\theta
+2\pi i)=\Gamma _{\hat{q}}[-\hat{\theta}-\frac{\tau }{2}]\Gamma _{\hat{q}}[1-%
\hat{\theta}+\frac{\tau }{2}].%
\end{array}%
\end{equation*}%
Hence, we notice that the solution of (\ref{c1}) and (\ref{c3}) is by no
means unique and there exist additional ones to the one presented in \cite%
{Z4} in form of a discrete integral representation. The function in the
numerator $\rho _{n,i}(\theta )$ may be selected by the requirement that we
would like to obtain the corresponding quantities of the sine-Gordon model
in the trigonometric limit and $\rho _{d,j}(\theta )$ by discarding
solutions which yield undesired poles. We end up with the choice $\rho
_{4,4}(\theta )=\rho _{n,4}(\theta )/\rho _{d,4}(\theta )$, which after
substitution into (\ref{ansatz}) yields. 
\begin{eqnarray}
c(\theta ) &=&\frac{\Gamma _{\hat{q}}[1+\frac{\tau }{2}]\Gamma _{\hat{q}}[-%
\frac{\tau }{2}]}{\Gamma _{\hat{q}}[1+\frac{\lambda }{2}]\Gamma _{\hat{q}}[-%
\frac{\lambda }{2}]}\prod\limits_{k=1}^{\infty }\frac{\Gamma _{\hat{q}}[\hat{%
\theta}-k\lambda ]\Gamma _{\hat{q}}[1+\hat{\theta}-(k-1)\lambda ]}{\Gamma _{%
\hat{q}}[-\hat{\theta}-k\lambda ]\Gamma _{\hat{q}}[1-\hat{\theta}%
-(k-1)\lambda ]}\quad ~  \notag \\
&&\times \frac{\Gamma _{\hat{q}}[-\hat{\theta}-(k-\frac{1}{2})\lambda
]\Gamma _{\hat{q}}[1-\hat{\theta}-(k-\frac{3}{2})\lambda ]\Gamma _{\hat{q}%
}[1+\hat{\theta}-k\lambda +\frac{\tau }{2}]}{\Gamma _{\hat{q}}[\hat{\theta}%
-(k+\frac{1}{2})\lambda ]\Gamma _{\hat{q}}[1+\hat{\theta}-(k-\frac{1}{2}%
)\lambda ]\Gamma _{\hat{q}}[1-\hat{\theta}-(k-1)\lambda +\frac{\tau }{2}]}~~
\label{cc} \\
&&\times \frac{\Gamma _{\hat{q}}[\hat{\theta}-k\lambda -\frac{\tau }{2}%
]\Gamma _{\hat{q}}[-\hat{\theta}-(k-\frac{1}{2})\lambda -\frac{\tau }{2}%
]\Gamma _{\hat{q}}[1-\hat{\theta}-(k-\frac{1}{2})\lambda +\frac{\tau }{2}]}{%
\Gamma _{\hat{q}}[-\hat{\theta}-(k-1)\lambda -\frac{\tau }{2}]\Gamma _{\hat{q%
}}[\hat{\theta}-(k-\frac{1}{2})\lambda -\frac{\tau }{2}]\Gamma _{\hat{q}}[1+%
\hat{\theta}-(k-\frac{1}{2})\lambda +\frac{\tau }{2}]}.  \notag
\end{eqnarray}

We want to conclude this section with a general remark on the method
provided to solve the functional relations (\ref{c1}) and (\ref{c3}). Of
course we could have started right away with the ansatz (\ref{ansatz})
instead of introducing the affine Weyl group in the first place. However,
this formulation automatically supplies a certain systematic. To illustrate
this further we provide another example of some important functional
relations occurring in the context of integrable models, that is Watson's
equations, see e.g.~\cite{SG}, for the minimal form factors 
\begin{equation}
F_{\text{min}}(\theta +i\pi )=F_{\text{min}}(i\pi -\theta )\qquad \text{%
and\qquad }F_{\text{min}}(\theta )=F_{\text{min}}(-\theta )S(\theta )~.
\label{Wat}
\end{equation}%
Making here an ansatz by taking the ratio of right cosets in which $N_{1}$
is even and odd 
\begin{equation}
F_{\text{min}}\left( \theta \right) =\kappa \frac{\left\{ \sigma
^{2n}I,N_{1}\right\} \left\{ \sigma ^{2n}\sigma _{0},N_{1}\right\} }{\left\{
\sigma ^{2n+1}I,N_{1}\right\} \left\{ \sigma ^{2n}\sigma _{1},N_{1}\right\} }%
=\kappa \prod\limits_{k=0}^{\infty }\frac{\sigma ^{2k-2}\rho \left( \theta
\right) \sigma ^{2k}\sigma _{0}\rho \left( \theta \right) }{\sigma ^{2k-1}%
\bar{\rho}\left( \theta \right) \sigma ^{2k}\sigma _{1}\bar{\rho}\left(
\theta \right) },
\end{equation}%
we see that the first equation in (\ref{Wat}) is solved by construction
whereas the second requires that $S(\theta )=\rho \left( \theta \right)
\sigma _{1}\bar{\rho}\left( \theta \right) /\sigma _{0}\rho \left( \theta
\right) \sigma \bar{\rho}\left( \theta \right) $. Having a concrete
expression for $S(\theta )$ one can now easily determine $\rho $ and $\bar{%
\rho}$. For more complicated functional relations one may use groups of
higher rank.

\section{Soliton-antisoliton sector}

Let us now discuss in more detail the elliptic sine-Gordon model. Scattering
amplitudes are obtainable in general from the computation of matrix
elements, but it is also well established that in an integrable ($\equiv $%
factorizable) theory they may be derived equivalently by analyzing the
Zamolodchikov algebra. Its associativity corresponds to the Yang-Baxter
equations. Its generators are thought of as particle creation operators and
therefore internal symmetries of the model are respected by this algebra.
Considering a theory with two particles which are conjugate to each other,
say a soliton $Z$ and an antisoliton $\bar{Z}$, one may demand a $\mathbb{Z}%
_{4}$-symmetry, that is one requires invariance under $Z\rightarrow \exp
(i\pi /2)Z$, $\bar{Z}\rightarrow \exp (-i\pi /2)\bar{Z}$. The most general
version of the Zamolodchikov algebra respecting this symmetry then reads 
\cite{Z4} 
\begin{eqnarray}
Z(\theta _{1})Z(\theta _{2}) &=&a(\theta _{12})Z(\theta _{2})Z(\theta
_{1})+d(\theta _{12})\bar{Z}(\theta _{2})\bar{Z}(\theta _{1})~,  \label{Z1}
\\
Z(\theta _{1})\bar{Z}(\theta _{2}) &=&b(\theta _{12})\bar{Z}(\theta
_{2})Z(\theta _{1})+c(\theta _{12})Z(\theta _{2})\bar{Z}(\theta _{1})~,
\label{Z2}
\end{eqnarray}
with rapidity difference $\theta _{12}=\theta _{1}-\theta _{2}$. The charge
conjugated relations also hold, that is $Z\leftrightarrow \bar{Z}$. In
comparison with the more extensively studied sine-Gordon model the
difference is the occurrence of the amplitude $d$ in (\ref{Z1}), i.e.~the
possibility that two solitons change into two antisolitons and vice versa.
Invoking the Yang-Baxter equations, crossing and unitarity one finds the
following solutions for the amplitudes

\begin{eqnarray}
a(\theta ) &=&\Phi (\theta )\prod\limits_{k=0}^{\infty }\frac{\Gamma _{\hat{q%
}^{2}}[\hat{\theta}-(k+1)\lambda ]\Gamma _{\hat{q}^{2}}[1+\hat{\theta}%
-k\lambda ]\Gamma _{\hat{q}^{2}}[-\hat{\theta}-\frac{1+2k}{2}\lambda ]\Gamma
_{\hat{q}^{2}}[1-\hat{\theta}-\frac{1+2k}{2}\lambda ]}{\Gamma _{\hat{q}%
^{2}}[-\hat{\theta}-(k+1)\lambda ]\Gamma _{\hat{q}^{2}}[1-\hat{\theta}%
-k\lambda ]\Gamma _{\hat{q}^{2}}[\hat{\theta}-\frac{1+2k}{2}\lambda ]\Gamma
_{\hat{q}^{2}}[1+\hat{\theta}-\frac{1+2k}{2}\lambda ]}\quad \quad  \label{a}
\\
b(\theta ) &=&-\frac{\func{sn}(i\theta /\nu )}{\func{sn}(i\theta /\nu +\pi
/\nu )}a(\theta )=\hat{b}(\theta )a(\theta ),  \label{b} \\
c(\theta ) &=&\frac{\func{sn}(\pi /\nu )}{\func{sn}(i\theta /\nu +\pi /\nu )}%
a(\theta )=\hat{c}(\theta )a(\theta ), \\
d(\theta ) &=&-\sqrt{\ell }\func{sn}(i\theta /\nu )\func{sn}(\pi /\nu
)a(\theta )~=\hat{d}(\theta )a(\theta ),  \label{d} \\
\Phi (\theta ) &=&\frac{\Gamma _{\hat{q}}[1+\frac{\tau }{2}]\Gamma _{\hat{q}%
}[-\frac{\tau }{2}]\Gamma _{\hat{q}}[1-\hat{\theta}+\frac{\lambda }{2}+\frac{%
\tau }{2}]\Gamma _{\hat{q}}[\hat{\theta}-\frac{\lambda }{2}-\frac{\tau }{2}]%
}{\Gamma _{\hat{q}}[1+\hat{\theta}+\frac{\tau }{2}]\Gamma _{\hat{q}}[-\hat{%
\theta}-\frac{\tau }{2}]\Gamma _{\hat{q}}[1+\frac{\lambda }{2}+\frac{\tau }{2%
}]\Gamma _{\hat{q}}[-\frac{\lambda }{2}-\frac{\tau }{2}]}~.
\end{eqnarray}
Here we use as a common factor $a(\theta )$ rather than $c(\theta )$ for
which we have explained in the previous section how to solve the key
functional equations (\ref{c1}) and (\ref{c3}). We also used relation (\ref%
{11}) to simplify (\ref{cc}) and refer the reader to \cite{Z4} for the
details on how to relate the different amplitudes to each other.

Our solutions (\ref{a})-(\ref{d}) differ in form from the one presented in 
\cite{Z4}, where a discretized version of an integral representation was
provided. Instead we derived here directly an infinite product
representation in the previous section, which is a natural generalization of
a very common version, of the form (\ref{gen}), used in the context of the
sine-Gordon model. One of the advantages of our formulation is that it
exhibits very explicitly the singularity structure. Furthermore, it is
easier to handle under shifts of the argument as in integral representations
such shifts will often be prohibited by convergence requirements.

\subsection{Singularity structure}

In order to extract the singularity structure for the amplitudes (\ref{a})-(%
\ref{d}) we recall the relations (\ref{pG}), (\ref{zJ}) and (\ref{pJ}),
which suffice to read off the poles and zeros of $a(\theta ),b(\theta
),c(\theta )$ and $d(\theta )$. We depict them most conveniently in a figure.

\FIGURE{\epsfig{file=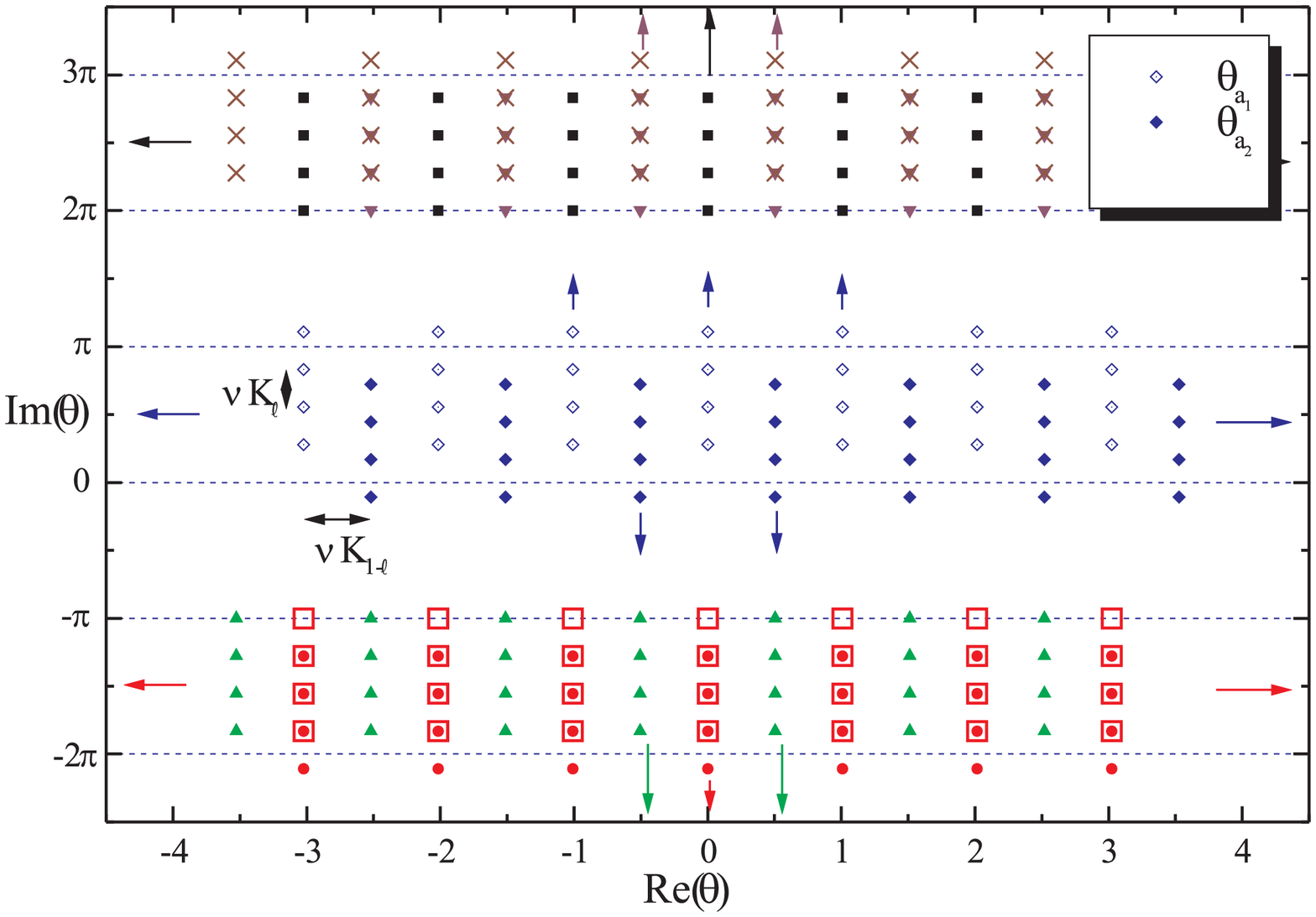,width=15cm} 
       \caption{Poles of  the soliton-soliton amplitude 
       $a(\theta) = b(i \pi -\theta) $ for the concrete values 
       $\nu=0.25$, $\ell=0.344$,  that is $\nu K_{1-\ell} \sim 0.504$  and
              $\nu K_{\ell} \sim 0.435$. The arrows indicate in which directions the
       ``strings" of poles are extended for growing integer values $n$ or $m$.
            }        \label{figure1}}

Explicitly, we just report here the poles which have potentially a chance to
be situated inside the physical sheet, i.e.~$0<\func{Im}\theta <\pi $. We
take $l,m\in \mathbb{Z},n\in \mathbb{N}$ and associate always two sets of
poles $\theta _{a_{1},p}^{nm}$ and $\theta _{a_{2},p}^{nm}$ to $a(\theta )$, 
$\theta _{b_{1},p}^{nm}$ and $\theta _{b_{2},p}^{nm}$ to $b(\theta )$ etc. 
\begin{equation}
\begin{array}{ll}
\theta _{a_{1},p}^{nm}=2m\nu K_{1-\ell }+i2n\nu K_{\ell }, & \theta
_{a_{2},p}^{nm}=(2m+1)\nu K_{1-\ell }+i(\pi -2n\nu K_{\ell }), \\ 
\theta _{b_{1},p}^{lm}=2m\nu K_{1-\ell }+i(\pi -2l\nu K_{\ell }),\qquad  & 
\theta _{b_{2},p}^{lm}=(2m+1)\nu K_{1-\ell }+i2l\nu K_{\ell }, \\ 
\theta _{c_{1},p}^{lm}=2m\nu K_{1-\ell }+i2l\nu K_{\ell }, & \theta
_{c_{2},p}^{lm}=2m\nu K_{1-\ell }+i(\pi -2l\nu K_{\ell }), \\ 
\theta _{d_{1},p}^{lm}=(2m+1)\nu K_{1-\ell }+i2l\nu K_{\ell }, & \theta
_{d_{2},p}^{lm}=(2m+1)\nu K_{1-\ell }+i(\pi -2n\nu K_{\ell }).%
\end{array}
\label{poles}
\end{equation}%
One readily sees from (\ref{poles}) that if one restricts the parameter $\nu
\geq \pi /2K_{\ell }$ all poles move out of the physical sheet into the
non-physical one, where they can be interpreted in principle as resonances,
i.e.~unstable particles. This was already stated in \cite{Z4} where the
choice $\nu \geq \pi /2K_{\ell }$ was made in order to avoid the occurrence
of tachyonic states. In fact, even in that regime there are Tachyons
present, since the poles in the non-physical sheet with negative real part
can not be explained on the basis of the Breit-Wigner formula. Their
occurence can be avoided by an additional breaking of parity (see discussion
in \cite{C30F}). The restriction on the parameters makes the model somewhat
unattractive as this limitation eliminates the analogue of the entire
breather sector which is present in the sine-Gordon model, such also that in
the trigonometric limit one only obtains the soliton-antisoliton sector of
that model, instead of a theory with a richer particle content. For this
reason, the arguments outlined in the introduction and the fact that the
constraint does not yield any Tachyon free theory anyhow, we relax here the
restriction on $\nu $. The the poles 
\begin{equation}
\theta _{b_{1},p}^{n0}=\theta _{c_{2},p}^{n0}\qquad \quad ~~\text{for~\ }%
0<n<n_{\max }=[\pi /2\nu K_{\ell }],n\in \mathbb{N}  \label{bp}
\end{equation}%
are located on the imaginary axis in the physical sheet and are therefore
candidates for the analogue of the n$^{th}$-breather bound states in the
sine-Gordon model. We indicate here the integer part of $x$ by $[x]$. In
other words, there are at most $n_{\max }-1$ breathers for fixed $\nu $ and $%
\ell $. The price one pays for the occurrence of these new particles in the $%
\mathbb{Z}_{4}$-model is that one unavoidably also introduces additional
Tachyons into the model as the poles always emerge in \textquotedblleft
strings\textquotedblright . It remains to be established whether the poles (%
\ref{bp}) may really be associated to a breather type behaviour.

\section{Soliton-breather amplitudes}

Once the solitonic sector of a theory is constructed there is a well defined
bootstrap procedure proposed by Karowski and Thun \cite{Karo, KT}, which
allows to complete the theory and to construct also the breather sector.
There are numerous solutions known for a non-trivial soliton sector, since
for instance for all affine Toda field theories with purely imaginary
coupling constant the Yang-Baxter equations can be solved by representations
of the corresponding quantum group \cite{nond1, nond2, Bassi:1999ua}.
Nonetheless, the completion of the models has not been carried out for very
many cases \cite{nond3, nond4, nond5}. In reverse, one should note that
there exist many scalar theories which do not possess a known solitonic
counterpart. One reason for that is, that the procedure \cite{Karo, KT} is
not reversible and one can in general not construct the solitonic sector
from the breather sector alone. Drawing a loose analogy to group theory one
may think of the breather sector as a subgroup, which of course does not
contain the information of a larger group in which it might be embedded. It
is very desirable to complete the picture, tie up the loose ends and
construct the respective missing sectors.

To start the construction, one first of all has to find creation operators
for the particles corresponding to the bound state poles in the soliton
sector. We presume here that the particles related to the poles (\ref{bp})
are breathers and borrow some intuition from the classical theory,
exploiting the fact that a breather is an oscillatory object made out of a
superposition of a soliton and an antisoliton. For the sine-Gordon model
this prescription was used in \cite{KT} to define the n$^{\text{th}}$%
-breather particle creation operator. Even though we do not have a classical
counterpart for the elliptic sine-Gordon model, we follow the same approach
here and define the auxiliary state \ 
\begin{equation}
Z_{n}(\theta _{1},\theta _{2}):=\frac{1}{\sqrt{2}}\left[ Z(\theta _{1})\bar{Z%
}(\theta _{2})+(-1)^{n}\bar{Z}(\theta _{1})Z(\theta _{2})\right] ~.
\label{breather}
\end{equation}%
Obviously, this state has properties of the classical sine-Gordon breather
being chargeless and having parity $(-1)^{n}$. Choosing thereafter the
rapidities such that the state (\ref{breather}) is on-shell, we can speak of
a breather bound state 
\begin{equation}
\lim_{(p_{1}+p_{2})^{2}\rightarrow m_{b_{n}}^{2}}Z_{n}(\theta _{1},\theta
_{2})\equiv \lim_{\theta _{12}\rightarrow \theta +\theta
_{12}^{b_{n}}}Z_{n}(\theta _{1},\theta _{2})=Z_{n}(\theta )~.
\label{onshell}
\end{equation}%
Here $\theta _{12}^{b_{n}}$ is the fusing angle related to the poles in the
soliton-antisoliton scattering amplitudes. To simplify notation we
abbreviate in what follows $a_{ij}=a(\theta _{ij})$, $b_{ij}=b(\theta _{ij})$%
, etc. With the help of the braiding relations (\ref{Z1}) and (\ref{Z2}) we
compute the scattering amplitude between a soliton and the auxiliary
breather states (\ref{breather}) 
\begin{eqnarray}
Z_{n}(\theta _{1},\theta _{2})Z(\theta _{3}) &=&\frac{1}{\sqrt{2}}\left[
a_{13}b_{23}Z(\theta _{3})Z(\theta _{1})\bar{Z}(\theta
_{2})+(-1)^{n}b_{13}a_{23}Z(\theta _{3})\bar{Z}(\theta _{1})Z(\theta
_{2})\right.   \notag \\
&&+c_{13}c_{23}Z(\theta _{3})\bar{Z}(\theta _{1})Z(\theta
_{2})+(-1)^{n}d_{13}d_{23}Z(\theta _{3})Z(\theta _{1})\bar{Z}(\theta _{2}) 
\notag \\
&&+b_{13}c_{23}\bar{Z}(\theta _{3})Z(\theta _{1})Z(\theta
_{2})+(-1)^{n}c_{13}a_{23}\bar{Z}(\theta _{3})Z(\theta _{1})Z(\theta _{2})
\label{sb} \\
&&+\left. d_{13}b_{23}\bar{Z}(\theta _{3})\bar{Z}(\theta _{1})\bar{Z}(\theta
_{2})+(-1)^{n}a_{13}d_{23}\bar{Z}(\theta _{3})\bar{Z}(\theta _{1})\bar{Z}%
(\theta _{2})\right] ~.  \notag
\end{eqnarray}%
Going now on-shell in the sense as defined in (\ref{onshell}), splitting the
fusing angle into $\theta _{1}=\theta _{0}-$~$\theta _{21}^{n}/2$, $\theta
_{2}=\theta _{0}+$~$\theta _{21}^{n}/2$ and denoting $\theta _{03}=\theta $
the following identities can be shown to hold 
\begin{eqnarray}
b_{13}c_{23}+(-1)^{n}c_{13}a_{23} &=&0\qquad \qquad \qquad \qquad \quad
\qquad \text{for ~~}\theta _{21}^{n}=i(\pi -2n\nu K_{\ell })  \label{id1} \\
d_{13}b_{23}+(-1)^{n}a_{13}d_{23} &=&0\qquad \qquad \qquad \qquad \qquad
\quad \text{for ~~}\theta _{21}^{n}=i(\pi -2n\nu K_{\ell }) \\
b_{13}a_{23}+(-1)^{n}c_{13}c_{23} &=&a_{13}b_{23}+(-1)^{n}d_{13}d_{23}\qquad 
\text{for ~~}\theta _{21}^{n}=i(\pi -2n\nu K_{\ell })~.  \label{id3}
\end{eqnarray}%
Therefore the braiding relation (\ref{sb}) reduces to a diagonal scattering
process, i.e.~there is no backscattering amplitude. We obtain after some
algebra 
\begin{equation}
Z_{n}(\theta _{0})Z(\theta _{3})=S_{b_{n}s}(\theta )Z(\theta
_{3})Z_{n}(\theta _{0})~,  \label{ns}
\end{equation}%
with 
\begin{eqnarray}
S_{b_{n}s}(\theta ) &=&a_{13}b_{23}+(-1)^{n}d_{13}d_{23} \\
&=&\frac{\func{sn}(i\theta /\nu -\pi /2\nu +nK_{\ell })}{\func{sn}(i\theta
/\nu +\pi /2\nu +nK_{\ell })}[\ell {\func{sn}}^{2}\frac{\pi }{\nu }{\func{sn}%
}^{2}\left( \frac{i\theta }{\nu }+\frac{\pi }{2\nu }+nK_{\ell }\right)
-1]a_{13}a_{23}\quad   \label{sbs}
\end{eqnarray}%
where 
\begin{eqnarray}
&&a_{13}a_{23}=\Phi _{13}\Phi _{23}\frac{\Gamma _{\hat{q}^{2}}[1+\hat{\theta}%
+\frac{\lambda }{4}-\frac{n}{2}]\Gamma _{\hat{q}^{2}}[-\hat{\theta}-\frac{%
\lambda }{4}-\frac{n}{2}]\Gamma _{\hat{q}^{2}}[-\hat{\theta}+\frac{\lambda }{%
4}+\frac{n}{2}]\Gamma _{\hat{q}^{2}}[\hat{\theta}+\frac{\lambda }{4}-\frac{n%
}{2}]}{\Gamma _{\hat{q}^{2}}[1-\hat{\theta}+\frac{\lambda }{4}-\frac{n}{2}%
]\Gamma _{\hat{q}^{2}}[\hat{\theta}-\frac{\lambda }{4}-\frac{n}{2}]\Gamma _{%
\hat{q}^{2}}[\hat{\theta}+\frac{\lambda }{4}+\frac{n}{2}]\Gamma _{\hat{q}%
^{2}}[-\hat{\theta}+\frac{\lambda }{4}-\frac{n}{2}]}  \notag \\
&&\times \prod\limits_{k=0}^{\infty }\frac{[\hat{\theta}-\frac{n}{2}+\frac{%
\lambda }{4}-k\lambda ]_{\hat{q}^{2}}[-\hat{\theta}+\frac{n}{2}-\frac{%
\lambda }{4}-k\lambda ]_{\hat{q}^{2}}[\hat{\theta}+\frac{n}{2}+\frac{\lambda 
}{4}-k\lambda ]_{\hat{q}^{2}}[-\hat{\theta}-\frac{n}{2}-\frac{\lambda }{4}%
-k\lambda ]_{\hat{q}^{2}}}{[-\hat{\theta}-\frac{n}{2}+\frac{\lambda }{4}%
-k\lambda ]_{\hat{q}^{2}}[\hat{\theta}+\frac{n}{2}-\frac{\lambda }{4}%
-k\lambda ]_{\hat{q}^{2}}[-\hat{\theta}+\frac{n}{2}+\frac{\lambda }{4}%
-k\lambda ]_{\hat{q}^{2}}[\hat{\theta}-\frac{n}{2}-\frac{\lambda }{4}%
-k\lambda ]_{\hat{q}^{2}}}  \notag \\
&&\times \prod\limits_{l=1}^{n-1}\frac{[\hat{\theta}-\frac{n}{2}+\frac{%
\lambda }{4}-k\lambda +l]_{\hat{q}^{2}}^{2}[-\hat{\theta}+\frac{n}{2}-\frac{%
\lambda }{4}-k\lambda -l]_{\hat{q}^{2}}^{2}}{[-\hat{\theta}-\frac{n}{2}+%
\frac{\lambda }{4}-k\lambda +l]_{\hat{q}^{2}}^{2}[\hat{\theta}+\frac{n}{2}-%
\frac{\lambda }{4}-k\lambda -l]_{\hat{q}^{2}}^{2}}~.  \label{aa}
\end{eqnarray}%
Similarly, we compute the braiding of $Z_{n}(\theta _{1},\theta _{2})\bar{Z}%
(\theta _{3})$ which leads us to 
\begin{equation}
S_{b_{n}\bar{s}}(\theta )=S_{b_{n}s}(\theta )~.  \label{ssb}
\end{equation}%
Denoting the anti-particle always by an overbar, we deduce from (\ref{ssb})
that the breathers are self-conjugate due to the general relation $S_{ab}=$ $%
S_{\bar{a}\bar{b}}$, that is $\bar{b}_{n}=b_{n}$.

The matrix $S_{b_{n}s}(\theta )$ contains various types of poles. i) simple
and double poles inside the physical sheet beyond the imaginary axis, which
are redundant from our point of view as they are of a tachyonic nature, ii)
double poles on the imaginary axis which can be explained by the usual
\textquotedblleft box diagrams\textquotedblright\ corresponding to the
Coleman-Thun mechanism \cite{CT}, iii) simple poles in the non-physical
sheet which can be interpreted as unstable particles (see e.g.~\cite{C30F}
and references therein) and iv) one simple pole on the imaginary axis inside
the physical sheet at $\theta =i\pi /2+in\nu K_{\ell }$ which is associated
to a soliton produced as an n$^{\text{th}}$-breathers-soliton bound state.

\section{Breather-breather amplitudes}

We now proceed similarly as in the previous section and compute the
scattering amplitudes of the breathers amongst themselves. As mentioned,
this is a very interesting sector as, under certain circumstances which will
be specified below, it usually closes independently from the remaining
sectors of the model. Similarly as in (\ref{ns}) we exchange now two of the
auxiliary states (\ref{breather}) 
\begin{equation}
Z_{n}(\theta _{1},\theta _{2})Z_{m}(\theta _{3},\theta
_{4})=S_{b_{n}b_{m}}(\theta _{1},\theta _{2},\theta _{3},\theta
_{4})Z_{m}(\theta _{3},\theta _{4})Z_{n}(\theta _{1},\theta _{2})~.
\end{equation}%
This is a somewhat cumbersome computation and we will not report here the
analogue expression of (\ref{sb}), since it involves 64 terms, each one of
them consisting of a product of four amplitudes and four creation operators.
As in the previous section, in the next step we have to go on-shell by
specifying the fusing angles as in (\ref{onshell}). We choose for this
purpose $\theta _{1}=\theta _{0}-$~$\theta _{21}^{n}/2$, $\theta _{2}=\theta
_{0}+$~$\theta _{21}^{n}/2$, $\theta _{3}=\theta _{0}^{\prime }-$~$\theta
_{43}^{m}/2$, $\theta _{4}=\theta _{0}^{\prime }+$~$\theta _{43}^{m}/2$ with 
$\theta _{43}^{m}=i(\pi -2m\nu K_{\ell })$, $\theta _{21}^{n}=i(\pi -2n\nu
K_{\ell })$, $\theta =\theta _{0}-\theta _{0}^{\prime }$ such that 
\begin{eqnarray}
\theta _{13} &=&\theta +i(n-m)\nu K_{\ell },\quad \theta _{14}=\theta -\pi
i+i(n+m)\nu K_{\ell },\quad \\
\theta _{24} &=&\theta +i(m-n)\nu K_{\ell },\quad \theta _{23}=\theta +\pi
i-i(n+m)\nu K_{\ell }.
\end{eqnarray}%
For this choice of the fusing angles there are several on-shell identities
of the type (\ref{id1})-(\ref{id3}). Extracting here the common factors of $%
a_{ij}$ we compute 
\begin{eqnarray}
&&\left( -1\right) ^{m}\,(\hat{b}_{24}\,\hat{c}_{13}\,\hat{c}_{14}+\hat{b}%
_{14}\,\hat{d}_{23}\,\hat{d}_{24})+\left( -1\right) ^{n}\,(\hat{b}_{13}\,%
\hat{c}_{14}\,\hat{c}_{24}+\,\hat{b}_{14}\,\hat{d}_{13}\,\hat{d}_{23}) 
\notag \\
&&+\hat{c}_{13}\,\hat{c}_{14}\,\hat{c}_{23}\,\hat{c}_{24}+\hat{b}_{14}\,\hat{%
b}_{23}+\left( -1\right) ^{m+n}(\,\hat{b}_{14}\,\hat{b}_{23}\,\hat{d}_{13}\,%
\hat{d}_{24}+\,\hat{b}_{13}\,\hat{b}_{24}\,\hat{c}_{14}\,\hat{c}_{23})=
\label{00} \\
&&\left( -1\right) ^{m}\,(\hat{b}_{23}\,\hat{d}_{13}\,\hat{d}_{14}+\,\hat{b}%
_{13}\,\hat{c}_{23}\,\hat{c}_{24})+\left( -1\right) ^{n}\,(\hat{b}_{24}\,%
\hat{c}_{13}\,\hat{c}_{23}+\,\hat{b}_{23}\,\hat{d}_{14}\,\hat{d}_{24}) 
\notag \\
&&+\hat{d}_{13}\,\hat{d}_{14}\,\hat{d}_{23}\,\hat{d}_{24}+\hat{b}_{13}\,\hat{%
b}_{24}+\left( -1\right) ^{m+n}\,(\hat{d}_{14}\,\hat{d}_{23}+\,\hat{c}_{13}\,%
\hat{c}_{24})  \notag
\end{eqnarray}%
and 
\begin{eqnarray}
&&\left( -1\right) ^{m}\,(\hat{b}_{13}\,\hat{b}_{24}\,\hat{d}_{14}+\,\hat{d}%
_{13}\,\hat{d}_{23}\,\hat{d}_{24})+\left( -1\right) ^{n}\,(\hat{c}_{13}\,%
\hat{c}_{24}\,\hat{d}_{14}+\,\hat{d}_{23})  \notag \\
&&+\hat{b}_{13}\,\hat{c}_{23}\,\hat{c}_{24}\,\hat{d}_{14}+\hat{b}_{23}\,\hat{%
d}_{13}+\left( -1\right) ^{m+n}\,(\hat{b}_{23}\,\hat{d}_{24}+\,\hat{b}_{24}\,%
\hat{c}_{13}\,\hat{c}_{23}\,\hat{d}_{14})=0~,  \label{01} \\
&&\left( -1\right) ^{m}\,(\hat{b}_{13}\,\hat{b}_{14}\,\hat{c}_{24}+\,\hat{c}%
_{14}\,\hat{d}_{13}\,\hat{d}_{23})+\left( -1\right) ^{n}(\,\hat{b}_{14}\,%
\hat{b}_{24}\,\hat{c}_{13}+\,\hat{c}_{14}\,\hat{d}_{23}\,\hat{d}_{24}{)} 
\notag \\
&&+\hat{b}_{23}\,\hat{c}_{14}\,\hat{d}_{13}\,\hat{d}_{24}+\hat{b}_{13}\,\hat{%
b}_{14}\,\hat{b}_{24}\,\hat{c}_{23}+\left( -1\right) ^{m+n}\,(\hat{b}_{23}\,%
\hat{c}_{14}+\,\hat{b}_{14}\,\hat{c}_{13}\,\hat{c}_{23}\,\hat{c}_{24})=0~.
\label{02}
\end{eqnarray}%
The relations (\ref{01}) and (\ref{02}) lead to a cancellation of the
backscattering terms in an analogous fashion as in (\ref{sb}). With (\ref{00}%
) we indeed end up with a diagonal scattering matrix 
\begin{equation}
Z_{n}(\theta _{0})Z_{m}(\theta _{0}^{\prime })=S_{b_{n}b_{m}}(\theta
)Z_{m}(\theta _{0}^{\prime })Z_{n}(\theta _{0})
\end{equation}%
where

\begin{eqnarray}
S_{b_{n}b_{m}}(\theta ) &=&a_{13}a_{14}\,a{_{23}}\,\,a{_{24}}\left[ \hat{c}{%
_{13}}\,\hat{c}{_{14}}\,\hat{c}{_{23}}\,\hat{c}{_{24}}+\hat{b}_{14}\,\hat{b}%
_{23}{+(-1)}^{n+m}(\hat{b}{_{14}}\,\hat{b}{_{23}}\,\hat{d}{_{13}}\,\hat{d}{%
_{24}+}\hat{b}{_{13}}\,\hat{b}{_{24}}\,\hat{c}{_{14}}\,\hat{c}{_{23})}\right.
\notag \\
&&\left. +{(-1)}^{m}(\hat{b}{_{24}}\,\hat{c}{_{13}}\,\hat{c}{_{14}+}\hat{b}{%
_{14}}\hat{d}{_{23}}\,\hat{d}{_{24})}+{(-1)}^{n}(\hat{b}{_{14}}\,\hat{d}{%
_{13}}\,\hat{d}{_{23}}+\hat{b}{_{13}}\,\hat{c}{_{14}}\,\hat{c}{_{24})}\right]
\label{Sbb} \\
&=&\left[ 1-\ell {\func{sn}}^{2}\frac{\pi }{\nu }{\func{sn}}^{2}\left( \frac{%
i\theta }{\nu }+(n+m)K_{\ell }\right) \right] \left[ 1-\ell {\func{sn}}^{2}%
\frac{\pi }{\nu }{\func{sn}}^{2}\left( \frac{i\theta }{\nu }+(n+m)K_{\ell }+%
\frac{\pi }{\nu }\right) \right]  \notag \\
&&\times \frac{\func{sn}(i\theta /\nu -\pi /\nu +(n+m)K_{\ell })}{\func{sn}%
(i\theta /\nu +\pi /\nu +(n+m)K_{\ell })}a_{13}a_{14}\,a{_{23}}\,\,a{_{24}}
\label{Sbb2}
\end{eqnarray}%
and 
\begin{eqnarray}
a_{13}a_{14}\,a{_{23}}\,\,a{_{24}} &{=}&{\Phi }_{13}{\Phi }_{14}{\Phi }_{23}{%
\Phi }_{24}\frac{\Gamma _{\hat{q}^{2}}(1+{\frac{m}{2}}+{\frac{n}{2}}+\hat{%
\theta}+{\frac{\lambda }{2}})\,\Gamma _{\hat{q}^{2}}({\frac{-m}{2}}-{\frac{n%
}{2}}-\hat{\theta}-{\frac{\lambda }{2}})}{\Gamma _{\hat{q}^{2}}(1+{\frac{m}{2%
}}+{\frac{n}{2}}-\hat{\theta}+{\frac{\lambda }{2}})\Gamma _{\hat{q}^{2}}({%
\frac{-m}{2}}-{\frac{n}{2}}+\hat{\theta}-{\frac{\,\lambda }{2}})}
\label{aaaa} \\
&&\times \prod\limits_{k=1}^{\infty }\prod\limits_{l=1}^{n-1}\frac{[{\frac{m%
}{2}}+{\frac{n}{2}-l}-\hat{\theta}-k\,\lambda +\lambda ]_{\hat{q}^{2}}[{%
\frac{-m}{2}}-{\frac{n}{2}+l}+\hat{\theta}-k\,\lambda ]_{\hat{q}^{2}}}{[{%
\frac{m}{2}}+{\frac{n}{2}-l}+\hat{\theta}-k\,\lambda +\lambda ]_{\hat{q}%
^{2}}[{\frac{-m}{2}}-{\frac{n}{2}+l}-\hat{\theta}-k\,\lambda ]_{\hat{q}^{2}}}
\notag \\
&&\times \prod\limits_{k=0}^{\infty }\prod\limits_{l=1}^{n-1}\frac{[{\frac{m%
}{2}}+{\frac{n}{2}-l}+\hat{\theta}-{\frac{\lambda }{2}}-k\,\lambda ]_{\hat{q}%
^{2}}[-{\frac{m}{2}}-{\frac{n}{2}+l}-\hat{\theta}-{\frac{\lambda }{2}}%
-k\,\lambda ]_{\hat{q}^{2}}}{[{\frac{m}{2}}+{\frac{n}{2}-l}-\hat{\theta}-{%
\frac{\lambda }{2}}-k\,\lambda ]_{\hat{q}^{2}}[-{\frac{m}{2}}-{\frac{n}{2}+l}%
+\hat{\theta}-{\frac{\lambda }{2}}-k\,\lambda ]_{\hat{q}^{2}}}  \notag \\
&&\times \prod\limits_{k=1}^{\infty }\prod\limits_{l=0}^{m-1}\frac{[{\frac{m%
}{2}}+{\frac{n}{2}-l}-\hat{\theta}-k\,\lambda +\lambda ]_{\hat{q}^{2}}[-{%
\frac{m}{2}}-{\frac{n}{2}+l}+\hat{\theta}-k\,\lambda ]_{\hat{q}^{2}}}{[{%
\frac{m}{2}}+{\frac{n}{2}-l}+\hat{\theta}-k\,\lambda +\lambda ]_{\hat{q}%
^{2}}[-{\frac{m}{2}}-{\frac{n}{2}+l}-\hat{\theta}-k\,\lambda ]_{\hat{q}^{2}}}
\notag \\
&&\times \prod\limits_{k=0}^{\infty }\prod\limits_{l=0}^{m-1}\frac{[{\frac{m%
}{2}}+{\frac{n}{2}-l}+\hat{\theta}-{\frac{\lambda }{2}}-k\,\lambda ]_{\hat{q}%
^{2}}[-{\frac{m}{2}}-{\frac{n}{2}+l}-\hat{\theta}-{\frac{\lambda }{2}}%
-k\,\lambda ]_{\hat{q}^{2}}}{[{\frac{m}{2}}+{\frac{n}{2}-l}-\hat{\theta}-{%
\frac{\lambda }{2}}-k\,\lambda ]_{\hat{q}^{2}}[-{\frac{m}{2}}-{\frac{n}{2}+l}%
+\hat{\theta}-{\frac{\lambda }{2}}-k\,\lambda ]_{\hat{q}^{2}}}.  \notag
\end{eqnarray}%
The latter expression (\ref{aaaa}) is tailored to make contact to the
expressions in the literature corresponding to the trigonometric limit.

The matrix $S_{b_{n}b_{m}}(\theta )$ also exhibits several types of poles.
i) simple and double poles inside the physical sheet beyond the imaginary
axis, ii) double poles located on the imaginary axis, iii) simple poles in
the non-physical sheet and iv) one simple pole on the imaginary axis inside
the physical sheet at $\theta =\theta _{b}=i\nu (n+m)K_{\ell }$ which is
related to the fusing process of two breathers $b_{n}+b_{m}\rightarrow
b_{n+m}$. To be really sure that this pole admits such an interpretation, we
have to establish that the imaginary part of the residue is strictly
positive, i.e. 
\begin{equation}
-i\lim_{\theta \rightarrow \theta _{b}}(\theta -\theta
_{b})S_{b_{n}b_{m}}(\theta )>0~.  \label{sheap}
\end{equation}%
Since the scattering process is parity invariant, we have $%
S_{b_{n}b_{m}}=S_{b_{m}b_{n}}$, such that we can choose without loss of
generality $n\geq m$. With this choice we compute 
\begin{eqnarray}
&&\limfunc{Res}_{\theta \rightarrow \theta _{b}}S_{b_{n}b_{m}}(\theta
)=i(-1)^{n+m}\frac{2K_{1-\ell }\nu }{\pi }\sinh \left[ \pi \frac{K_{\ell }}{%
K_{1-\ell }}(n+m)\right] \left( 1-\ell {\func{sn}}^{4}\frac{\pi }{\nu }%
\right) \hat{q}^{-2(n+m)(n+m+\lambda +1)}  \notag \\
&&\times {\Phi }\left[ i\pi \right] {\Phi }\left[ 2in\nu K_{\ell }\right] {%
\Phi }\left[ 2in\nu K_{\ell }\right] {\Phi }\left[ 2i(n+m)\nu K_{\ell }-i\pi %
\right] \prod\limits_{l=1}^{n-1}\frac{[n+m-l]_{\hat{q}^{2}}}{[-l]_{\hat{q}%
^{2}}}\prod\limits_{l=1}^{m-1}\frac{[n+m-l]_{\hat{q}^{2}}}{[-l]_{\hat{q}^{2}}%
}  \notag \\
&&\times \prod\limits_{k=1}^{\infty }\left( \frac{[n-k\,\lambda ]_{\hat{q}%
^{2}}[-n-k\,\lambda ]_{\hat{q}^{2}}[n+\frac{\lambda }{2}-k\,\lambda ]_{\hat{q%
}^{2}}[-n+\frac{\lambda }{2}-k\,\lambda ]_{\hat{q}^{2}}[n+m-k\,\lambda ]_{%
\hat{q}^{2}}}{[m-k\,\lambda ]_{\hat{q}^{2}}[-m-k\,\lambda ]_{\hat{q}^{2}}[m+%
\frac{\lambda }{2}-k\,\lambda ]_{\hat{q}^{2}}[-m+\frac{\lambda }{2}%
-k\,\lambda ]_{\hat{q}^{2}}[n+m+\frac{\lambda }{2}-k\,\lambda ]_{\hat{q}^{2}}%
}\right.  \notag \\
&&\times \frac{\lbrack -n-m-k\,\lambda ]_{\hat{q}^{2}}[\frac{\lambda }{2}%
-k\,\lambda ]_{\hat{q}^{2}}^{2}}{[-n-m+\frac{\lambda }{2}-k\,\lambda ]_{\hat{%
q}^{2}}[-k\,\lambda ]_{\hat{q}^{2}}^{2}}\prod\limits_{l=1}^{m-1}\frac{%
[n+m-l-k\,\lambda ]_{\hat{q}^{2}}^{2}[l-n-m-k\,\lambda ]_{\hat{q}^{2}}^{2}}{%
[-l-k\,\lambda ]_{\hat{q}^{2}}^{2}[l-k\,\lambda ]_{\hat{q}^{2}}^{2}}  \notag
\\
&&\times \left. \frac{\lbrack \frac{\lambda }{2}+n+m-l-k\,\lambda ]_{\hat{q}%
^{2}}^{2}[\frac{\lambda }{2}-n-m+l-k\,\lambda ]_{\hat{q}^{2}}^{2}}{[\frac{%
\lambda }{2}-l-k\,\lambda ]_{\hat{q}^{2}}^{2}[\frac{\lambda }{2}%
+l-k\,\lambda ]_{\hat{q}^{2}}^{2}}\right)  \label{Wetter}
\end{eqnarray}%
The first line in (\ref{Wetter}) equals $i(-1)^{n+m}\kappa $ with $\kappa
\in \mathbb{R}^{+}$. Noting that the functions ${\Phi }$ with the above
arguments are positive real numbers, the second line in (\ref{Wetter}) is $%
(-1)^{n+m}\kappa ^{\prime }$ with $\kappa ^{\prime }\in \mathbb{R}^{+}$.
Recalling finally that $n+m<-\lambda /2$ we deduce that $\prod%
\nolimits_{k=1}^{\infty }(~)\in \mathbb{R}^{+}$ such that (\ref{sheap}) is
indeed satisfied.

Due to the factorizability of the theory this fusing process can be
associated in the usual fashion to a bootstrap equation. For consistency the
following equations have to be satisfied 
\begin{equation}
S_{lb_{n+m}}(\theta )=S_{lb_{n}}(\theta +i\nu mK_{\ell })S_{lb_{m}}(\theta
-i\nu nK_{\ell })\quad \text{for }l\in \{b_{k},s,\bar{s}\}~;k,m+n<n_{\max }~.
\label{boot}
\end{equation}%
With some algebra we verified (\ref{boot}) for the amplitudes derived above (%
\ref{sbs}) and (\ref{Sbb2}).

\section{Reductions of the $\mathbb{Z}_{4}$-model}

The elliptic sine-Gordon model can be considered as a master theory, which
contains many other theories as submodels. By choosing various specific
values for the two free parameters of the model $\ell $ and $\nu $, one
obtains these different types of theories. For the different choices we have
the following interrelations 
\begin{equation}
\begin{array}{ccc}
\fbox{elliptic sine-Gordon} & \overset{1/\nu \rightarrow 2nK_{\ell }/\pi
+2imK_{1-\ell }/\pi }{-------\longrightarrow } & \fbox{elliptic $%
D_{n+1}^{(1)}$-affine Toda theory} \\ 
\begin{array}{c}
| \\ 
| \\ 
\ell \rightarrow 0 \\ 
| \\ 
\downarrow%
\end{array}
& 
\begin{array}{ccc}
&  & \swarrow \\ 
& 
\begin{array}{c}
m\neq 0,\ell \rightarrow 0 \\ 
\downarrow \\ 
\fbox{free theory} \\ 
\uparrow \\ 
1/\nu \rightarrow i\infty%
\end{array}
&  \\ 
\nearrow &  & 
\end{array}
& 
\begin{array}{c}
| \\ 
| \\ 
m=0,\ell \rightarrow 0 \\ 
| \\ 
\downarrow%
\end{array}
\\ 
\fbox{sine-Gordon} & \underset{1/\nu \rightarrow n}{-------\longrightarrow }
& \fbox{minimal $D_{n+1}^{(1)}$-affine Toda theory}%
\end{array}
\label{diagram}
\end{equation}
Let us discuss this schematic diagram in more detail:

\subsection{Trigonometric limits}

As was already stated in \cite{Z4}, when carrying out one of the
trigonometric limit $\ell \rightarrow 0$ for the $\mathbb{Z}_{4}$-model for
the amplitudes (\ref{a})-(\ref{d}), one recovers the soliton sector of the
sine-Gordon model. In our formulation this can be seen directly, as we only
have to use the relations (\ref{gqn}) and (\ref{ts}) for the $\Gamma _{\hat{q%
}}$- and $\func{sn}$-functions and note that $\lim_{\ell \rightarrow 0}\Phi
=1$. We have employed here the somewhat generalized conventions of the
sine-Gordon model formulation in \cite{SG}. To make contact with the
infinite product representation as presented in the literature (see
equations (2.16) and appendix C in \cite{SG}), we also have to use the
identity 
\begin{equation}
\prod\limits_{k=0}^{\infty }\frac{\Gamma ^{2}(\frac{2k}{\nu }+\alpha )}{%
\Gamma (\frac{2k}{\nu }+\alpha +\gamma )\Gamma (\frac{2k}{\nu }+\alpha
-\gamma )}=\prod\limits_{k=0}^{\infty }\frac{\Gamma ^{2}(\frac{k\nu }{2}+%
\frac{\alpha \nu }{2})}{\Gamma (\frac{k\nu }{2}+\frac{\alpha \nu }{2}+\frac{%
\gamma \nu }{2})\Gamma (\frac{k\nu }{2}+\frac{\alpha \nu }{2}-\frac{\gamma
\nu }{2})}~,
\end{equation}%
after the limit $\ell \rightarrow 0$ is performed. As we demonstrated above,
when relaxing the constraint $\nu \geq \pi /2K_{\ell }$, the $\mathbb{Z}_{4}$%
- model exhibits also a breather sector. It is easily seen that the
corresponding amplitudes can also be obtained in that limit. The expressions
for the soliton-breather amplitude (\ref{sbs}) and (\ref{aa}) are tailored
in such a way that we obtain the corresponding amplitude in the sine-Gordon
model (see equation (20) in \cite{KT}) upon the use of (\ref{gqn}), (\ref%
{limn}) and 
\begin{equation}
\frac{\sin (x\pi /\lambda )}{\sin (y\pi /\lambda )}=\prod\limits_{k=1}^{%
\infty }\frac{(x-k\lambda +\lambda )(-x-k\lambda )}{(y-k\lambda +\lambda
)(-y-k\lambda )}.  \label{sin}
\end{equation}%
Similarly, we recover the breather-breather amplitude of the sine-Gordon
model (see equation (22) in \cite{KT}) using in addition to (\ref{sin}) also 
\begin{equation}
\frac{\cos (x\pi /\lambda )}{\cos (y\pi /\lambda )}=\prod\limits_{k=0}^{%
\infty }\frac{(y-k\lambda -\frac{\lambda }{2})(-y-k\lambda -\frac{\lambda }{2%
})}{(x-k\lambda -\frac{\lambda }{2})(-x-k\lambda -\frac{\lambda }{2})}.
\end{equation}

There is of course the other trigonometric limit $\ell \rightarrow 1$, which
one could in principle compute by exploiting the relations between the $q$
and $\hat{q}$-deformed quantities mentioned in section 2. However, just by
considering the pole structure (\ref{poles}) and recalling (\ref{kk}), we
see that there are no poles left inside the physical sheet which could
produce a bound state, such that this theory will only possess a soliton
sector. We will not consider this case here.

Whereas these limits served essentially only as a consistency check, the
next one will lead to a new type of theory.

\subsection{Diagonal limit}

For the sine-Gordon model it is well known \cite{KF,KKF} that in the limit $%
\nu \rightarrow 1/n$ the backscattering amplitude vanishes and one obtains a
diagonal S-matrix which can be identified with a minimal $D_{n+1}^{(1)}$-
affine Toda field theory. We observe, that there is an analogue to this
behaviour in the $\mathbb{Z}_{4}$-model as also in this case the
backscattering amplitudes vanish in the limit 
\begin{equation}
1/\nu \rightarrow 1/\nu _{n,m}=(2nK_{\ell }+i2mK_{1-\ell })/\pi ~.
\label{nnm}
\end{equation}%
We find 
\begin{eqnarray}
\lim_{\nu \rightarrow \nu _{n,m}}c(\theta ) &=&0~\qquad ~~\text{for~}n,m\in 
\mathbb{Z},\qquad \\
\lim_{\nu \rightarrow \nu _{n,m}}d(\theta ) &=&0~~\qquad ~~\text{for~}n,m\in 
\mathbb{Z}.
\end{eqnarray}%
For the remaining amplitudes (\ref{a}) and (\ref{b}) in the soliton sector
we compute for this limit 
\begin{eqnarray}
\lim_{\nu \rightarrow \nu _{n,m}}b(\theta ) &=&(-1)^{n+1}a_{d}(\theta
)~\qquad ~~\text{for~}n,m\in \mathbb{Z}, \\
\lim_{\nu \rightarrow \nu _{n,m}}a(\theta ) &=&a_{d}(\theta )~\qquad ~~~~\
~~~~~~~~\text{for~}n,m\in \mathbb{Z}~.  \label{111}
\end{eqnarray}%
where 
\begin{eqnarray}
&&\quad a_{d}(\theta )=\prod\limits_{k=0}^{\infty }\prod\limits_{l=0}^{n-1}%
\frac{[-\frac{\theta }{i\pi }(n+m\tau )+(2k+1)n+l]_{\hat{q}^{2}}[\frac{%
\theta }{i\pi }(n+m\tau )+(2k+1)n-l]_{\hat{q}^{2}}}{[\frac{\theta }{i\pi }%
(n+m\tau )+(2k+1)n+l]_{\hat{q}^{2}}[-\frac{\theta }{i\pi }(n+m\tau
)+(2k+1)n-l]_{\hat{q}^{2}}}~  \label{add} \\
&&~\times \frac{\Gamma _{\hat{q}}[1+\frac{\tau }{2}]\Gamma _{\hat{q}}[-\frac{%
\tau }{2}]\Gamma _{\hat{q}}[1-n+(\frac{1}{2}-m)\tau +\frac{\theta }{i\pi }%
(n+m\tau )]\Gamma _{\hat{q}}[n+(m-\frac{1}{2})\tau -\frac{\theta }{i\pi }%
(n+m\tau )]~}{\Gamma _{\hat{q}}[1-n+(\frac{1}{2}-m)\tau ]\Gamma _{\hat{q}%
}[n+(m-\frac{1}{2})\tau ]\Gamma _{\hat{q}}[1+\frac{\tau }{2}-\frac{\theta }{%
i\pi }(n+m\tau )]\Gamma _{\hat{q}}[-\frac{\tau }{2}+\frac{\theta }{i\pi }%
(n+m\tau )]}  \notag
\end{eqnarray}%
This is obtained when replacing in (\ref{a}) 
\begin{equation}
\lim_{\nu \rightarrow \nu _{n,m}}\hat{\theta}\rightarrow -\frac{\theta }{%
i\pi }(n+m\tau )\quad \text{and}\quad \lim_{\nu \rightarrow \nu
_{n,m}}\lambda \rightarrow -2(n+m\tau ),
\end{equation}%
and exploiting the property (\ref{ola1}) thereafter. Of course one may also
carry out the limit (\ref{nnm}) for the soliton-breather and for the
breather-breather amplitude. We do not report those expressions here as they
are quite obvious, unlike in (\ref{add}), where several cancellations could
be carried out and the infinite product in the $\Gamma _{\hat{q}}$ could be
turned into products in $[~]_{\hat{q}^{2}}$. The scattering matrix obtained
in this way belongs to a new type of theory. One should mention that the
case $m\neq 0$, which complexifies the coupling constant is most likely only
of a formal nature, but we expect the case $m=0$ to be a meaningful theory.
However, this requires more analysis especially since there are additional
Tachyons present.

Let us now carry out the limit $\ell \rightarrow 0$ and verify that the
above mentioned diagram (\ref{diagram}) is indeed commutative. For the cases 
$m=0,n\geq 3$ we reproduce the entire scattering matrix of the minimal $%
D_{n+1}^{(1)}$-affine Toda field theory. In particular when carrying out the
limit in (\ref{add}) we find, upon the use of (\ref{sin}) 
\begin{eqnarray}
\lim_{\hat{q}\rightarrow 1,m=0}a_{d}(\theta ) &=&\prod\limits_{k=0}^{\infty
}\prod\limits_{l=0}^{n-1}\frac{[n(-x+2k+1)+l][n(x+2k+1)-l]}{%
[n(x+2k+1)+l][n(-x+2k+1)-l]}  \label{gn} \\
&=&\prod\limits_{l=0}^{n-1}\frac{\sinh \frac{1}{2}\left( \theta +\frac{i\pi l%
}{n}\right) }{\sinh \frac{1}{2}\left( \theta -\frac{i\pi l}{n}\right) }, 
\notag
\end{eqnarray}%
which coincides with the $S_{n+1,n+1}=S_{n,n}$-amplitudes in the minimal $%
D_{n+1}^{(1)}$-affine Toda field theory. For the cases $m=0,n=1,2$ \ we
reproduce the minimal $A_{1}^{(1)}\otimes A_{1}^{(1)},A_{3}^{(1)}$-affine
Toda field theories. We remark that these correspondences hold up to a
change of statistics, that is some amplitudes are only recovered up to a
factor of $-1$, which changes bosonic to fermionic statistics or vice versa.
These facts make it natural to call the above theories elliptic versions of
the associated limiting theory. As we mentioned above, taking the breather
sector alone constitutes a consistent theory in itself. For $\hat{q}=1,m=0$
we have minimal $A_{2n-2}^{(2)}\subset D_{n+1}^{(1)}$-affine Toda field
theory, a property which transcends also into the elliptic version. A final
comment is related to the center of the diagram. We find $\lim_{\hat{q}%
\rightarrow 1,m\neq 0}a_{d}(\theta )=1$ simply due to the property (\ref%
{limnn}).

\section{Conclusion}

We demonstrated that when one relaxes the constraint on the coupling
constants, one can construct a consistent breather sector for the elliptic
sine-Gordon model. The scattering of the breathers amongst themselves and
with the soliton sector satisfies a bootstrap equation related to the fusing
of two breathers to a third. For the formulation of the scattering
amplitudes we used as natural objects q-deformed functions. Roughly speaking
one replaces in the amplitudes in the soliton-antisoliton sector the
infinite products of Euler's gamma functions by a q-deformed version and in
the infinite product in the breather sector integers by q-deformed ones.
This tailors the models automatically in a form which allows to carry out
various limits.   Instead of carrying out the bootstrap analysis one could
alternatively take a spin chain as a starting point and use a method based
on the algebraic Bethe ansatz, pursued for instance in \cite{Doikou:1999xt},
to compute the breather S-matrix amplitudes. It would be interesting to
compare that approach with our findings.

In the diagonal limit we obtain an interesting new theory, which can be
viewed as an elliptic generalization of the minimal $D_{n+1}^{(1)}$-affine
Toda field theory. In \cite{Inf} we proposed a procedure which also lead to
elliptic generalizations of theories whose scattering amplitudes can be
expressed in terms of trigonometric functions. The theories obtained in that
fashion were, however, of a quite different nature. The procedure in \cite%
{Inf} works strictly on the principle that there are no redundant poles
present in the amplitudes, such that as a difference the string of tachyonic
states which was encountered here was confined to the non-physical sheet
where they can be viewed as unstable particles. Further investigation is
needed to clarify more the interpretation of these models and in particular
to establish whether they posses a meaningful conformal limit \cite{Prep}.

From a mathematical point of view it will be interesting to generalize the
method of section 3 to affine Weyl groups of higher rank.

\textbf{Acknowledgments: }We are grateful to the Deutsche
Forschungsgemeinschaft (Sfb288), for financial support. This work is
supported by the EU network ``EUCLID, \emph{Integrable models and
applications: from strings to condensed matter}'', HPRN-CT-2002-00325.

\bibliographystyle{phreport}
\bibliography{Ref}

\begin{thebibliography}{10}

\bibitem{Z4}
A.~Zamolodchikov,
\newblock $Z_4$-symmetric factorised S-matrix in two space-time dimensions,
\newblock Comm. Math. Phys. {\bf 69}, 165 (1979).

\bibitem{Bax}
R.~Baxter,
\newblock Partition function of the eight vertex lattice model,
\newblock Ann. Phys. {\bf 70}, 193 (1972).

\bibitem{Shankar:1981jq}
R.~Shankar,
\newblock A simple derivation of the Baxter model free energy,
\newblock Phys. Rev. Lett. {\bf 47}, 1177 (1981).

\bibitem{Shankar:1982kz}
R.~Shankar,
\newblock On the solution of some vertex models using factorizable S-matrices,
\newblock J. Statist. Phys. {\bf 29}, 649 (1982).

\bibitem{Sen:1999nx}
A.~Sen and B.~Zwiebach,
\newblock Tachyon condensation in string field theory,
\newblock JHEP {\bf 03}, 002 (2000).

\bibitem{Berkovits:2000hf}
N.~Berkovits, A.~Sen, and B.~Zwiebach,
\newblock Tachyon condensation in superstring field theory,
\newblock Nucl. Phys. {\bf B587}, 147 (2000).

\bibitem{Kutasov:2000qp}
D.~Kutasov, M.~Mari\~no, and G.~Moore,
\newblock Some exact results on tachyon condensation in string field theory,
\newblock JHEP {\bf 10}, 045 (2000).

\bibitem{Rastelli:2000hv}
L.~Rastelli, A.~Sen, and B.~Zwiebach,
\newblock String field theory around the tachyon vacuum,
\newblock Adv. Theor. Math. Phys. {\bf 5}, 353 (2002).

\bibitem{Harvey:2000na}
J.~Harvey, D.~Kutasov, and E.~Martinec,
\newblock On the relevance of tachyons,
\newblock hep-th/0003101  (2000).

\bibitem{Witten:2000nz}
E.~Witten,
\newblock Noncommutative tachyons and string field theory,
\newblock hep-th/0006071  (2000).

\bibitem{Adams:2001sv}
A.~Adams, J.~Polchinski, and E.~Silverstein,
\newblock Don't panic! Closed string tachyons in ALE space-times,
\newblock JHEP {\bf 10}, 029 (2001).

\bibitem{Andreev:2002sn}
O.~Andreev,
\newblock On tachyon condensation in string theory: World sheet computations,
\newblock Fortsch. Phys. {\bf 50}, 670 (2002).

\bibitem{Sen:2002qa}
A.~Sen,
\newblock Time and tachyon,
\newblock hep-th/0209122  (2002).

\bibitem{Buchel:2002tj}
A.~Buchel, P.~Langfelder, and J.~Walcher,
\newblock Does the tachyon matter?,
\newblock Annals Phys. {\bf 302}, 78 (2002).

\bibitem{Taylor:2002uv}
W.~Taylor,
\newblock Lectures on D-branes, tachyon condensation, and string field theory,
\newblock hep-th/0301094  (2002).

\bibitem{Sen:2002an}
A.~Sen,
\newblock Field theory of tachyon matter,
\newblock Mod. Phys. Lett. {\bf A17}, 1797 (2002).

\bibitem{Tomaschitz:1999tt}
R.~Tomaschitz,
\newblock Cosmic tachyon background radiation,
\newblock Int. J. Mod. Phys. {\bf A14}, 4275 (1999).

\bibitem{Choudhury:2002xu}
D.~Choudhury, D.~Ghoshal, D.~Jatkar, and S.~Panda,
\newblock On the cosmological relevance of the tachyon,
\newblock Phys. Lett. {\bf B544}, 231 (2002).

\bibitem{Shiu:2002qe}
G.~Shiu and I.~Wasserman,
\newblock Cosmological constraints on tachyon matter,
\newblock Phys. Lett. {\bf B541}, 6 (2002).

\bibitem{Ohta:2002ac}
K.~Ohta and T.~Yokono,
\newblock Gravitational approach to tachyon matter,
\newblock Phys. Rev. {\bf D66}, 125009 (2002).

\bibitem{Gibbons:2002md}
G.~Gibbons,
\newblock Cosmological evolution of the rolling tachyon,
\newblock Phys. Lett. {\bf B537}, 1 (2002).

\bibitem{Kofman:2002rh}
L.~Kofman and A.~Linde,
\newblock Problems with tachyon inflation,
\newblock JHEP {\bf 07}, 004 (2002).

\bibitem{Gibbons:2003gb}
G.~Gibbons,
\newblock Thoughts on tachyon cosmology,
\newblock hep-th/0301117  (2003).

\bibitem{Sami:2003qx}
M.~Sami, P.~Chingangbam, and T.~Qureshi,
\newblock Cosmological aspects of rolling tachyon,
\newblock hep-th/0301140  (2003).

\bibitem{Kim:2003qz}
C.~Kim, H.~Kim, Y.~Kim, and O.~Kwon,
\newblock Cosmology of rolling tachyon,
\newblock hep-th/0301142  (2003).

\bibitem{Karo}
M.~Karowski,
\newblock On the bound state problem in 1+1 dimensional field theories,
\newblock Nucl. Phys. {\bf B153}, 244 (1979).

\bibitem{KT}
M.~Karowski and H.~Thun,
\newblock Complete S-matrix of the massive Thirring model,
\newblock Nucl. Phys. {\bf B130}, 295 (1977).

\bibitem{Migo}
M.~M{\"u}ller,
\newblock Form factors in integrable quantum field theory, generalizations of
  the homogeneous sine-Gordon model,
\newblock Diplomarbeit Freie Universit{\"a}t Berlin, unpublished  (2002).

\bibitem{Coleman}
S.~Coleman,
\newblock Quantum sine-Gordon equation as the massive Thirring model,
\newblock Phys. Rev. {\bf D11}, 2088 (1975).

\bibitem{Elliptic}
K.~Chandrasekhan,
\newblock Elliptic Functions,
\newblock Springer Verlag Berlin  (1985).

\bibitem{Hum2}
J.~Humphreys,
\newblock Reflection Groups and Coxeter Groups,
\newblock Cambridge University Press  (1990).

\bibitem{SG}
H.~Babujian, A.~Fring, M.~Karowski, and A.~Zapletal,
\newblock Exact form factors in integrable quantum field theories: The
  sine-Gordon model,
\newblock Nucl. Phys. {\bf B538}, 535 (1999).

\bibitem{C30F}
O.~A. Castro-Alvaredo, J.~Drei{\ss}ig, and A.~Fring,
\newblock Integrable scattering theories with unstable particles,
\newblock hep-th/0211168 .

\bibitem{nond1}
T.~Nakatsu,
\newblock Quantum group approach to affine Toda field theory,
\newblock Nucl. Phys. {\bf B356}, 499 (1991).

\bibitem{nond2}
T.~J. Hollowood,
\newblock Solitons in affine Toda field theories,
\newblock Nucl. Phys. {\bf B384}, 523 (1992).

\bibitem{Bassi:1999ua}
Z.~S. Bassi and A.~LeClair,
\newblock The exact S-matrix for an $osp(2|2)$ disordered system,
\newblock Nucl. Phys. {\bf B578}, 577 (2000).

\bibitem{nond3}
G.~M. Gandenberger,
\newblock Exact S matrices for bound states of $a_2^{(1)}$ affine Toda
  solitons,
\newblock Nucl. Phys. {\bf B449}, 375 (1995).

\bibitem{nond4}
G.~M. Gandenberger and N.~J. MacKay,
\newblock Exact S matrices for $d_{(N+1)}^{(2)}$ affine Toda solitons and their
  bound states,
\newblock Nucl. Phys. {\bf B457}, 240 (1995).

\bibitem{nond5}
G.~M. Gandenberger, N.~J. MacKay, and G.~M.~T. Watts,
\newblock Twisted algebra R-matrices and S-matrices for $b_n^{(1)}$ affine Toda
  solitons and their bound states,
\newblock Nucl. Phys. {\bf B465}, 329 (1996).

\bibitem{CT}
S.~R. Coleman and H.~J. Thun,
\newblock On the prosaic origin of the double poles in the sine-Gordon
  S-matrix,
\newblock Commun. Math. Phys. {\bf 61}, 31 (1978).

\bibitem{KF}
V.~E. Korepin and L.~D. Faddeev,
\newblock Quantization of solitons,
\newblock Theor. Math. Phys. {\bf 25}, 1039 (1975).

\bibitem{KKF}
V.~E. Korepin, P.~P. Kulish, and L.~D. Faddeev,
\newblock Soliton quantization,
\newblock JETP Lett. {\bf 21}, 138 (1975).

\bibitem{Doikou:1999xt}
A.~Doikou and R.~I. Nepomechie,
\newblock Direct calculation of breather S matrices,
\newblock J. Phys. {\bf A32}, 3663 (1999).

\bibitem{Inf}
O.~A. Castro-Alvaredo and A.~Fring,
\newblock Constructing infinite particle spectra,
\newblock Phys. Rev. {\bf D64}, 085005 (2001).

\bibitem{Prep}
O.~A. Castro-Alvaredo and A.~Fring,
\newblock in preparation.

\end{thebibliography}

\end{document}